# Spatial analysis and prediction of COVID-19 spread in South Africa after lockdown

Mohammad Arashi [1], Andriette Bekker [2*], Mahdi Salehi [3], Sollie Millard [2], Barend Erasmus [4], Tanita Cronje [2], and Mohammad Golpaygani [5]

[1]Department of Statistics, Faculty of Mathematical Science, Shahrood University of Technology, Shahrood, Iran; m_arashi_stat@yahoo.com

[2]Department of Statistics, Faculty of Natural and Agricultural Science, University of Pretoria, Pretoria, South Africa; andriette.bekker@up.ac.za; sollie.millard@up.ac.za; tanita.cronje@up.ac.za

[3]Department of Mathematics and Statistics, Faculty of Basic Sciences, University of Neyshabur, Iran; salehi2sms@gmail.com

[4]Office of the Dean, Faculty of Natural and Agricultural Science, University of Pretoria, Pretoria, South Africa; barend.erasmus@up.ac.za

[5]Department of Statistics, Faculty of Sirjan School of Medical Sciences, Sirjan, Iran; golpa313@gmail.com

*Correspondence: andriette.bekker@up.ac.za;

**Abstract:** What is the impact of COVID-19 on South Africa? This paper envisages assisting researchers and decision-makers in battling the COVID-19 pandemic focusing on South Africa. This paper focuses on the spread of the disease by applying heatmap retrieval of hotspot areas and spatial analysis is carried out using the Moran index. For capturing spatial autocorrelation between the provinces of South Africa, the adjacent, as well as the geographical distance measures, are used as a weight matrix for both absolute and relative counts. Furthermore, generalized logistic growth curve modeling is used for the prediction of the COVID-19 spread. We expect this data-driven modeling to provide some insights into hotspot identification and timeous action controlling the spread of the virus.

## 1. Introduction

During December 2019, several cases of pneumonia of an unknown aetiology were reported in Wuhan, a city within the Hubei province of China ([1]). Within a week investigators found that the initial cases where all associated with a seafood market where live poultry and wild animals were being sold ([2]). Since then the disease has been registered, and become known, as the coronavirus or COVID-19 which is caused by the Severe Acute Respiratory Syndrome Coronavirus 2 (SARS-CoV2). This disease has shown that in early stages of infection, symptoms of severe acute respiratory infection can occur. These may include a cough, fever and shortness of breath ([3]). Some patients may then develop acute respiratory distress syndrome (ARDS) and other serious complications which may potentially lead to multiple organ failure ([1]). Since mid-December, COVID-19 has spread to all seven continents, increasing its prevalence throughout the entire world, and was declared a pandemic, by the World Health Organization (WHO) on the 11th of March ([4]). This rapid spread has been fuelled by the fact that the majority of infected people do not experience severe symptoms, which makes it more likely to remain mobile and hence infect others ([5]). The transmission primarily occurs through contact from person to person, coughing or sneezing and touching of contaminated surfaces ([6]).

On the 14th of February the first case of COVID-19 was reported in Africa, in the City of Cairo, by the Egyptian Ministry of Health and Population. The individual, who travelled between China and Cairo on a business trip was identified through contact screening ([7]). The first South African case was confirmed by The National Institute for Communicable Diseases (NICD) on the morning of the 5th of March 2020. The patient, a 38 year old male, was part of a group of 10 people, including his wife, who arrived back in South Africa on the 1st of March from Europe. Since then the number of infections and deaths have risen drastically. President Cyril Rhamaphosa was praised by the director-general of the WHO, Dr Tedros Adganom Ghemreyesus, for his leadership and approach to protecting South Africans during these trying times ([8]). The British Broadcasting Corporation (BBC) also commended President Cyril Ramaphosa for his leadership and for South Africa's "ruthlessly efficient" response to the coronavirus ([9]). On the 15th of March President Cyril Ramaphosa declared a national state of disaster, the terms of the Disaster Management Act which enable the focus to be put on preventing and reducing the risk of the virus spreading ([10]), and only a few days later on the 23rd of March the President declared a national lockdown commencing on midnight of the 26th of March.

In South Africa, these extreme measures are absolutely necessary, as the country contains a high risk population combined with low-income country characteristics. The main concerns, which are thought to escalate the spread of the coronavirus, are the large and densely populated areas and townships, including a high level of poverty and movement within these areas. Combined with existing epidemics such as the human immunodeficiency virus (HIV), tuberculosis (TB) and malaria, this might lead to an increase in morbidity and mortality. Since the wide spread of non-communicable diseases, such as chronic obstructive pulmonary disease (COPD), heart disease, hypertension and diabetes, in Africa are known risk factors for severe cases of COVID-19 these may also increase the death rate in these lower-income countries ([11]). As winter is approaching, overcrowded houses and the large immunocompromised population, will contribute to the increase in the number of COVID-19 cases rising ([12]).

To date, as reported by the Coronavirus disease 2019 [13], South Africa has the largest number of COVID-19 cases in Africa. Although it has been shown that South Africans are generally complying during the lockdown, by investigating insights from vehicle-tracking data, is was shown that vehicle activity dropped by 20% even before the lockdown and reduced by 75% after the lockdown was implemented ([14]). This decline in movement directly indicates the effect on the economy, with the closure of businesses like manufacturing, retail and restaurants, to name only a few. With numerous businesses no longer operating, many South Africans are no longer receiving an income. The Human Science Research Council also released a note on the mental health of South Africans, stating that the stage is already set for major mental health implications, and noting that that failure to put measures into place to mitigate the psychological impacts of quarantine, is likely to lead to an ineffective and slow economic recovery ([15]).

In this study, we make use of the Moran index to spatially identify the spread in South Africa with respect to the provinces for the COVID-19 infections. Finding these hotspots will provide insights in identifying, and assist in tracking, the COVID-19 spread. With this information South Africa will be better able to predict local outbreaks and develop public health policies to better manage and update medical procedures currently set in place. Since the strict lockdown (level 5) is phased out from the 1st of May (moved to level 4 lockdown), the location of these hotspots could assist in guiding the risk-adjusted strategy and the economic activity plan, set out by the South African government. It would thus be important to know where these hotspots are and if they are statistically significant. Further, a generalized logistic model of the growth trend will be employed to show the difference between the hotspot areas and the areas outside of it. With the continuing growth and development of COVID-19 in South Africa, this analysis might be helpful to guide political leaders and health authorities to manage the allocation of resources and prepare for future virus control.

The effect of COVID-19 is still in early stages in South Africa but different tendencies have already been observed when compared to the US and other European countries ([13]). Understanding these tendencies will be very important in guiding the fight against COVID-19 in South African as well as the rest of Africa. This is an initial study from which many other interesting studies will follow and it will be very important to continue with analyses as more cases are reported and more data becomes available. South Africa, with the most confirmed COVID-19 cases, will need to be the leader in guiding the fight against COVID-19 in Africa.

## 2. Study Area and Materials

South Africa, formally known as the Republic of South Africa, is situated at the southernmost tip of Africa and covers a surface area of 1 219 602 km. With a coastline stretching more than 3000km from the desert border of Namibia touching the Atlantic Ocean, around the tip of Africa to the northern bordered of Mozambique on the Indian Ocean side. South Africa shares common boundaries with Namibia, Botswana, Zimbabwe, Swaziland, with the Mountain Kingdom of Lesotho landlocked by SA. The Prince Edward and Marion islands lie some 1 920km south-east of Cape Town ([16]).

With a population of more than 59 million, South Africa is the world's 25'th most populated nation consisting of nine different provinces. South Africa has three designated capital cities; executive Pretoria, judicial Bloemfontein and legislative Cape Town. The largest city and main economic hub being Johannesburg, which is also the main entry point for visitors from other countries via OR Thambo International Airport ([16]).

The following timeline of the major interventions in South Africa for the COVID-19 outbreak and the statistics are shown in Figures 1 and 2 respectively.

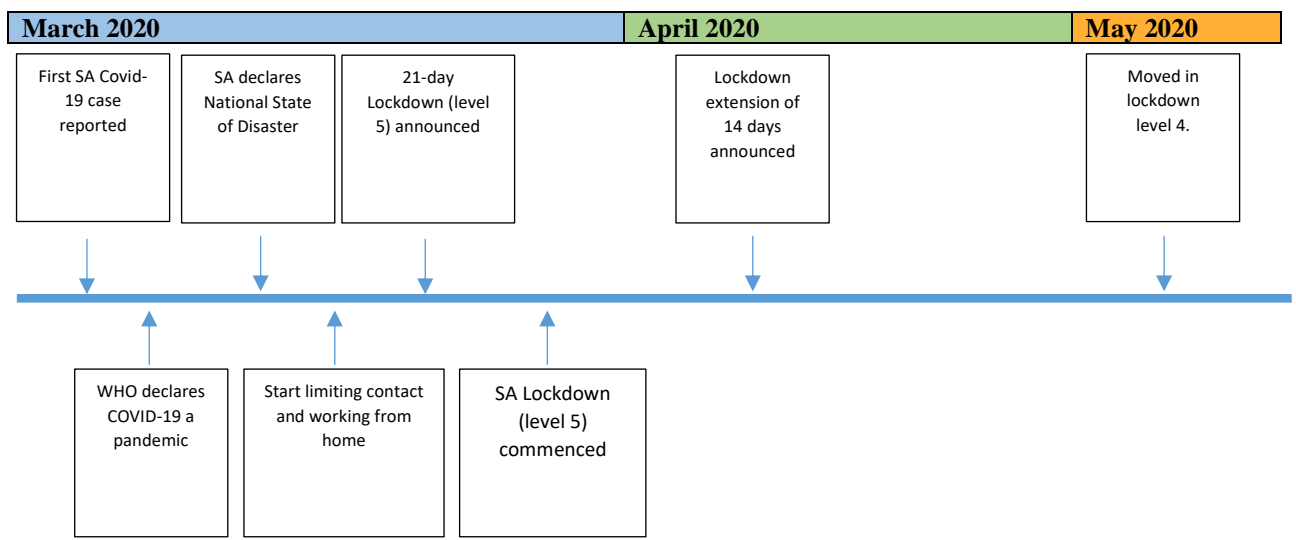

**Figure 1.** Timeline of major interventions.

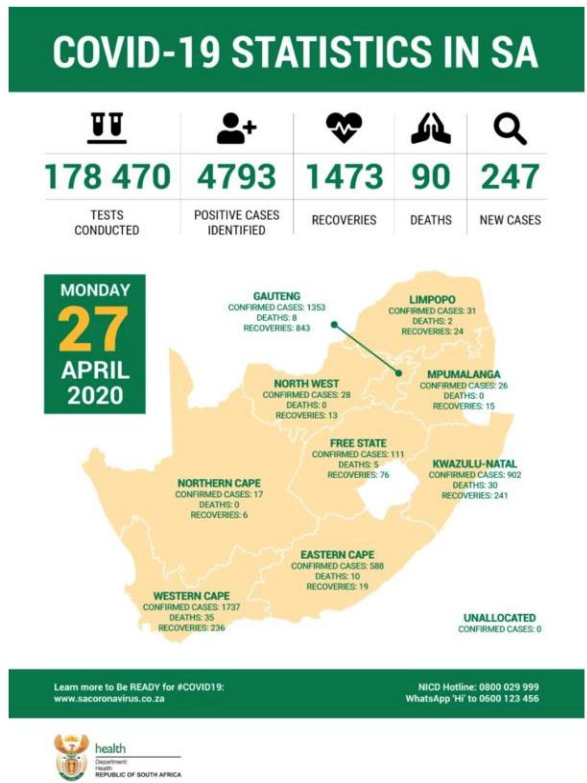

**Figure 2.** Covid-19 epidemic in South Africa.

The Data Science for Social Impact research (DSFSI) group at the University of Pretoria captures the COVID-19 data, number of cases, on national and provincial level. Missing values and anomalies in the provincial data are adjusted or imputed using data from the University of Cape Town COVID-19 dashboard. The demographic data used are the provincial population numbers published by. Number of recoveries and deaths used are those published by the NICD.

Distances between the provinces are determined using the main city from each of the provinces. The main city is the city most likely to be the highest risk for COVID-19 infection. In most cases, the main city is also the capital city of the province. The main cities are indicated in Table 1.

**Table 1.** Main cities in each of the provinces of South Africa

| Province | Main City |
|---|---|
| Eastern Cape (EC) | Port Elizabeth |
| Free State (FS) | Bloemfontein |
| Gauteng (GP) | Johannesburg |
| Kwazulu Natal (KZ) | Durban |
| Limpopo (LP) | Polokwane |
| Mpumalanga (MP) | Mbombela |
| North Cape (NC) | Kimberley |
| North West (NW) | Klerksdorp |
| Western Cape (WC) | Cape Town |

## 3. Methods

The spatial correlation between the 9 provinces, Northern Cape, Eastern Cape, Free State, Western Cape, Limpopo, North West, KwaZulu-Natal, Mpumalanga, and Gauteng in South Africa, we use the Moran's autocorrelation coefficient, also known as the Moran index (denoted by I) in geographic health science. Furthermore, we make use of generalized logistic function (GLF) for identifying an appropriate growth curve of COVID-19. Hence, this section is devoted to the definition of Moran index (Moran's I) as well as the GLF.

3.1. Spatial correlation coefficient

The Moran index, originally defined by [17], is a measure of spatial association or spatial autocorrelation which can be used to find spatial hotspots or clusters and is available in many software applications. This index has been defined as the measure of choice for scientists, specifically in environmental sciences, ecology and public health ([18]). Some other indices include the Getis' G index, Geary's C, local Ii and Gi, spatial scan statistics and Tango's C index ([19] and [18]). The Moran index has both a local and global representation. The global Moran's I is a global measure for spatial autocorrelation while the local Moran's I index examines the individual locations, enabling hotspots to be identified based on comparisons to the neighbouring locations ([19]). This local Moran's I has been successfully applied to hotspot identification for infection clusters such as those investigated by [20], who researched the bovine tuberculosis breakdowns (bTB )in Northern Irish cattle herds in order to access the spatial association in the number and prevalence of chronic bTB across Northern Ireland. Other areas where this index has been successfully applied and commonly used are diseases, mortality rates, environmental planning and environmental sciences. It's important to note that the result can be affected by the definition of the weight function, data transformation and existence of outliers ([19]).

Until now, not many COVID-19 related research has made use of the Moran Index and no research was found for South African specific cases. Some studies that include the use of this index are: 1) [21] explored the spatial epidemic dynamic of COVID-19 in mainland China in order to determine whether a spatial association of the COVID-19 infection existed; 2) [22] applied the Moran index to a spatial panel which showed that COVID-19 infection is spatially dependent and mainly spread from Hubei Province in Central Chine to neighbouring areas; 3) [23] used a global dataset of COVID-19 cases as well as a global climate database and investigated how climate parameters could contribute to the growth rate of COVID-19 cases while simultaneously controlling for potential confounding effects using spatial analysis; 4) [24] used data on all mobile phone users to examine the impact of the Coronavirus outbreak under the Swedish mild recommendations and restrictions regime on individual mobility and if the changes in geographical mobility vary over different socio-economic strata and 5) [25] investigated the influence of spatial proximities and travel patterns from Italy on the further spread of the SARS-CoV-2 around the globe.

This index is an extension of the Pearson's product-moment correlation coefficient for spatial pattern recognition. Observations in close proximity are more likely to be similar than those far apart ([26-27]). In order to formulate the Moran index for our purpose, assume we have $d$ provinces and the pair $(x_i, x_j)$ is for the attribute (variable) $x$ in provinces $i, j = 1, \cdots, d$, respectively. Then, the spatial weight $w_{ij}$ quantifies the level of closeness between $x_i$ and $x_j$ and the Moran index is defined by

$$\mathcal{I} = \frac{d}{S_o} \times \frac{\sum_{i=1}^{d}\sum_{j=1}^{d} w_{ij}(x_i - \bar{x})(x_j - \bar{x})}{\sum_{i=1}^{d}(x_i - \bar{x})^2}, \qquad (1)$$

where $\bar{x} = n^{-1}\sum_{i=1}^{n} x_i$ and $S_o = \sum_{i=1}^{n}\sum_{j=1}^{n} w_{ij}; i \neq j$.

The Moran's $\mathcal{I}$ takes value on $[-1,1]$ and $\mathcal{I} = 0$ shows no spatial correlation between the provinces for the underlying attribute. According to [28], there are two ways to identify the weights. In our context, we identify the $(i,j)$-th element of the weight matrix $W$, from taxonomic level classification viewpoint, as

$$w_{ij} = \begin{cases} 1 & \text{if the provinces i and j are connected} \\ 0 & \text{otherwise} \end{cases}$$

Using the phylogenetic tree classification (geographical distance), we assign the weights following

$$w_{ij} = \begin{cases} \frac{1}{D_{ij}^{\alpha}} & \text{if } D_{ij} \leq \lambda \\ 0 & \text{if } D_{ij} > \lambda \end{cases}$$

where $D_{ij}$ is the distance between the province centre $i$ and province centre $j$, $\lambda$ is a distance threshold, and $\alpha$ is a power level parameter. See [29] for more detail and comparison between different weights.

3.2. Modeling population growth

In this section, we predict some attributes via logistic growth curve modeling. The logistic function/curve is commonly used for dynamic modeling in many branches of science including chemistry, physics, material science, forestry, disease progression, sociology, etc. For our purpose and generality, we follow the Richards' differential equation (RDE) due to [30] given by

$$\frac{dP(t)}{dt} = \frac{1}{\alpha}\left(1 - \left(\frac{P(t)}{K}\right)^{\nu}\right) P(t)$$

with initial condition $P(t_o) = P_o$, $K$ is the carrying capacity, the maximum capacity or total population here, $\alpha, \nu > 0$ to obtain the generalized logistic curve (GLC)

$$P(t) = \frac{K}{\left(1 + Qe^{-\frac{\nu(t-t_o)}{\alpha}}\right)^{\frac{1}{\nu}}} \qquad (2)$$

with $Q = \left(\frac{K}{P_o}\right)^{\nu} - 1$.

The typical logistic curve which is widely used in modeling, is the special of the GLC for $\nu = 1$. Further, the Gompertz curve can be obtained for the limiting case $\nu \to 0^+$. See [31] for more details and applications of the GLC.

While only a few studies applied the logistic growth models to COVID-19 specific research questions, only one combined the model with the use of the Moran index to show that the infection is spatially dependent ([22]), with no studies for South African data. Some of these studies, which applied only the logistic growth model include; 1) [32] who uses the logistic growth equation to describe the process on a macroscopic level and 2) [33] who reviews the epidemic virus growth and decline curves in China using the phenomenological logistic growth model.

Mathematical description of biological growth (i.e. population models) is very important in many research disciplines. Amongst these population models, especially noteworthy are the clear analytical solutions of the generalized logistic functions ([34]).

## 4. Results and Discussions

In this section, we start off with a general inspection on the provincial distribution of the total confirmed and death cases given in Figures 3 and 4, respectively. From these figures it is observed that the heatmaps of confirmed and death cases agree, and therefore more confirmed cases are followed by more deaths. Furthermore, it is observed that the hotspots are Western Cape and Gauteng, with the former the highest risk of infection.

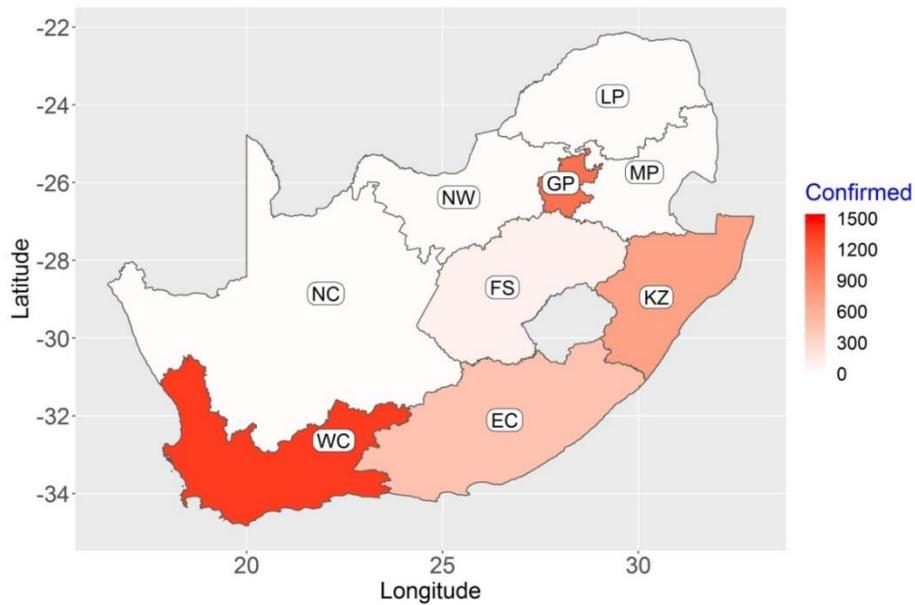

**Figure 3.** The heat map of the provincial distribution of total confirmed cases of COVID-19 up to April 25 in South Arica.

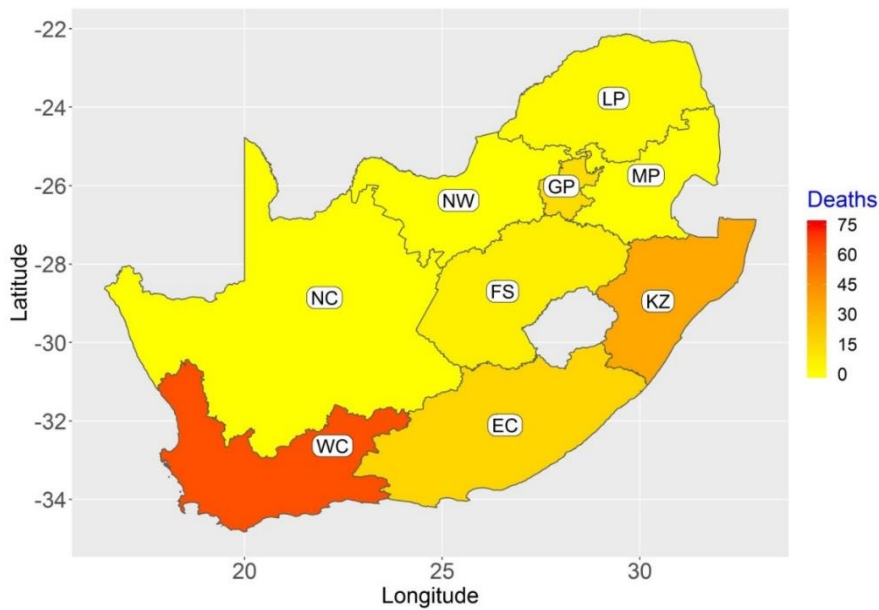

**Figure 4.** The heat map of the provincial distribution of total deaths of COVID-19 up to 25 April in South Arica.

In order to test the spatial autocorrelation of COVID-19 in South Africa, the interaction between provinces is estimated using Moran's I from March 21, 2020 to April 25, 2020 based on absolute counts by using Eq. (1) and the results are reported in Table 2.

Based on the adjacent 0-1 weight matrix, not all Moran coefficients are significant at the significance level of 5% (see the fifth column in Table 2), and the values are around the interval of [0, 1], since there is a positive correlation among the confirmed cases according to the geographical structure. Comparatively, no significant spatial correlation is tested out based on spatial geographic distance (the last column of Table 2), which indicates the spreading direction in South Africa is mainly based on adjacent areas to neighbors, and doesn't matter how far the distance to the infectious center. So main cities adjacent to are at higher risk.

**Table 2.** Moran's $I$ (observed), its expected value and the corresponding test for different days based on absolute counts.

| Date | Measurement by adjacency | | | | Measurement by geographical distance | | | |
|---|---|---|---|---|---|---|---|---|
| | Obs. | Exp. | sd | p-value | Obs. | Exp. | sd | p-value |
| 2020-03-21 | 0.0815 | -0.125 | 0.1354 | 0.1274 | -0.0864 | -0.125 | 0.1618 | 0.8117 |
| 2020-03-22 | 0.0847 | -0.125 | 0.1364 | 0.1240 | 0.0125 | -0.125 | 0.1624 | 0.3972 |
| 2020-03-23 | 0.0349 | -0.125 | 0.0751 | 0.0333 | -0.1974 | -0.125 | 0.1249 | 0.5619 |
| 2020-03-24 | 0.0401 | -0.125 | 0.0588 | 0.0050 | -0.2295 | -0.125 | 0.1174 | 0.3734 |
| 2020-03-25 | 0.1742 | -0.125 | 0.0712 | 0.0000 | 0.0743 | -0.125 | 0.1230 | 0.1051 |
| 2020-03-26 | -0.0350 | -0.125 | 0.1276 | 0.4805 | -0.2217 | -0.125 | 0.1564 | 0.5363 |
| 2020-03-27 | 0.0229 | -0.125 | 0.0807 | 0.0671 | -0.2602 | -0.125 | 0.1278 | 0.2901 |
| 2020-03-28 | -0.0203 | -0.125 | 0.0743 | 0.1589 | -0.0755 | -0.125 | 0.1245 | 0.6912 |
| 2020-03-29 | 0.0244 | -0.125 | 0.1316 | 0.2560 | -0.1666 | -0.125 | 0.1591 | 0.7937 |
| 2020-03-30 | 0.0341 | -0.125 | 0.0997 | 0.1105 | -0.2001 | -0.125 | 0.1385 | 0.5876 |
| 2020-03-31 | 0.1352 | -0.125 | 0.1078 | 0.0158 | -0.0815 | -0.125 | 0.1435 | 0.7617 |
| 2020-04-01 | 0.0000 | -0.125 | 0.1180 | 0.2892 | -0.2532 | -0.125 | 0.1500 | 0.3924 |
| 2020-04-02 | 0.0217 | -0.125 | 0.1472 | 0.3188 | -0.0544 | -0.125 | 0.1701 | 0.6783 |
| 2020-04-03 | 0.1654 | -0.125 | 0.1173 | 0.0133 | 0.0337 | -0.125 | 0.1495 | 0.2885 |
| 2020-04-04 | 0.1603 | -0.125 | 0.0946 | 0.0026 | 0.0815 | -0.125 | 0.1355 | 0.1273 |
| 2020-04-05 | 0.1575 | -0.125 | 0.1396 | 0.0430 | -0.0218 | -0.125 | 0.1647 | 0.5309 |
| 2020-04-06 | -0.0075 | -0.125 | 0.1502 | 0.4339 | -0.1234 | -0.125 | 0.1723 | 0.9925 |
| 2020-04-07 | 0.0212 | -0.125 | 0.1317 | 0.2670 | -0.1393 | -0.125 | 0.1592 | 0.9285 |
| 2020-04-08 | 0.0103 | -0.125 | 0.1329 | 0.3084 | -0.1500 | -0.125 | 0.1600 | 0.8756 |
| 2020-04-09 | 0.2414 | -0.125 | 0.1371 | 0.0075 | -0.1203 | -0.125 | 0.1629 | 0.9769 |
| 2020-04-10 | 0.1764 | -0.125 | 0.1388 | 0.0299 | -0.0055 | -0.125 | 0.1641 | 0.4666 |
| 2020-04-11 | -0.0469 | -0.125 | 0.1180 | 0.5082 | -0.2087 | -0.125 | 0.1500 | 0.5767 |
| 2020-04-12 | 0.0279 | -0.125 | 0.1446 | 0.2905 | -0.2188 | -0.125 | 0.1683 | 0.5772 |
| 2020-04-13 | 0.1084 | -0.125 | 0.1539 | 0.1294 | -0.1436 | -0.125 | 0.1750 | 0.9155 |
| 2020-04-14 | 0.2426 | -0.125 | 0.1083 | 0.0007 | -0.1369 | -0.125 | 0.1438 | 0.9340 |
| 2020-04-15 | 0.1445 | -0.125 | 0.1531 | 0.0783 | -0.2729 | -0.125 | 0.1744 | 0.3964 |

| Date | | | | | | | | |
|---|---|---|---|---|---|---|---|---|
| 2020-04-16 | -0.0052 | -0.125 | 0.1398 | 0.3917 | -0.3249 | -0.125 | 0.1649 | 0.2254 |
| 2020-04-17 | 0.0628 | -0.125 | 0.1548 | 0.2250 | -0.2004 | -0.125 | 0.1757 | 0.6676 |
| 2020-04-18 | 0.1563 | -0.125 | 0.1283 | 0.0283 | -0.0525 | -0.125 | 0.1569 | 0.6438 |
| 2020-04-19 | 0.0410 | -0.125 | 0.1397 | 0.2347 | -0.2705 | -0.125 | 0.1648 | 0.3772 |
| 2020-04-20 | 0.2683 | -0.125 | 0.0941 | 0.0000 | 0.0770 | -0.125 | 0.1352 | 0.1351 |
| 2020-04-21 | 0.2956 | -0.125 | 0.1298 | 0.0012 | 0.0097 | -0.125 | 0.1579 | 0.3935 |
| 2020-04-22 | 0.2187 | -0.125 | 0.1399 | 0.0140 | -0.0588 | -0.125 | 0.1649 | 0.6883 |
| 2020-04-23 | 0.3102 | -0.125 | 0.0735 | 0.0000 | 0.1166 | -0.125 | 0.1241 | 0.0515 |
| 2020-04-24 | 0.3758 | -0.125 | 0.1103 | 0.0000 | 0.0860 | -0.125 | 0.1450 | 0.1456 |
| 2020-04-25 | 0.2569 | -0.125 | 0.0579 | 0.0000 | 0.0994 | -0.125 | 0.1170 | 0.0552 |

To extend the analysis, we calculated the corresponding p-values of Moran's test given by Table 2 over the time, shown in Figure 5. Comparing the corresponding p-values of Moran's test, some deviation exist in the statistical timeliness in main cities of provinces from March 21, 2020 to April 25 so it is inevitable that maybe bias occurred in our results. From April 20 to 25, the spatial autocorrelation is significantly different in terms of adjacency to main cities. This means that the prevalence of COVID-19 varies in main cities.

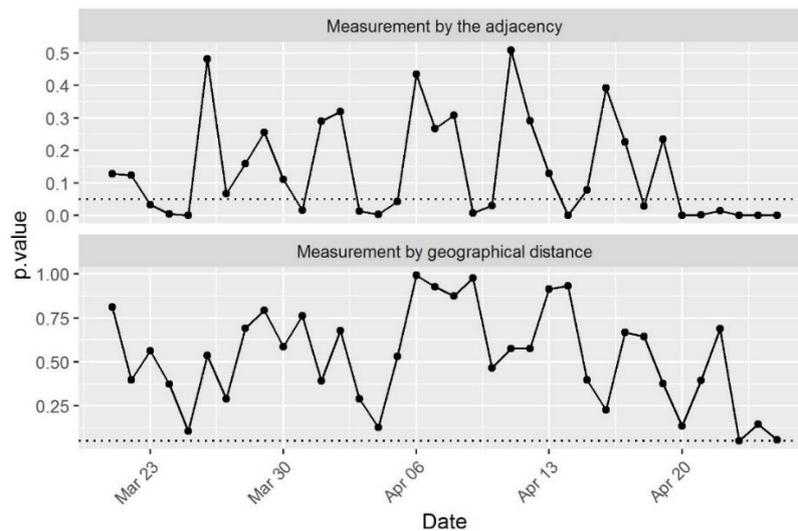

**Figure 5.** The corresponding p-values of Moran's test given by Table 2 over the time, on the basis of the absolute counts.

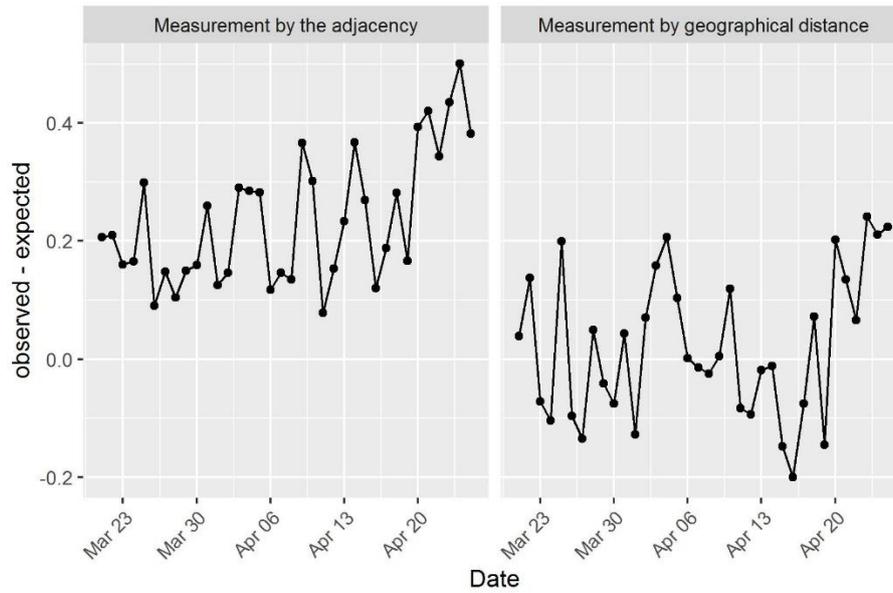

**Figure 6.** The difference between Moran's observed and expected values given by Table 2 over the time, on the basis of the absolute counts.

Additional analysis has been taken into account to validate the results based on the Moran index characteristics. The observed Moran index along with its expected value and standard error are tabulated in Table 3. The expected value of Moran index is -1/(N-1)=-1/8=-0.125 in our case. Using the adjacency weights, we obtain the same result as discussed based on Table 2. However, since the null hypothesis that there is no spatial autocorrelation between the provinces is not rejected for the geographical distance, we can argue there is no evidence of negative auto-correlation here, as with random data you would expect it to be a negative value more often than positive.

The impact of President Cyril Ramaphosa's decision in containing the outbreak by strict lockdown regulations is supported by the p-values in Figure 6 (see dates 20 April and onwards).

In addition, the spatial autocorrelation of COVID-19 in South Africa based on the relative counts, has been estimated using I from March 21, 2020 to April 25, 2020, and the results are reported in Table 3. In this table, for the adjacent 0-1 weights, more coefficients are significant at the level of 5% (see the fifth column in Table 4), with values in the interval [0, 1], since there is no positive correlation among the confirmed cases according to the geographical structure. Comparatively, very few significant spatial correlations are measured based on the spatial geographic distance (the last column of Table 3), which indicates that the spread direction in South Africa is mainly based on adjacent areas to neighbours, and less so on the distance to the infection centre. In conclusion, as it is identified in heatmaps Figures 3 and 4 the two main cities are at higher risk.

**Table 3.** Moran's $I$ (observed), its expected value and the corresponding test for different days based on relative counts.

| | Measurement by adjacency | | | | Measurement by geographical distance | | | |
|---|---|---|---|---|---|---|---|---|
| Date | Obs. | Exp. | sd | p-value | Obs. | Exp. | sd | p-value |
| 2020-03-21 | 0.1997 | -0.125 | 0.0912 | 0.0004 | 0.0321 | -0.125 | 0.1335 | 0.2393 |
| 2020-03-22 | 0.1476 | -0.125 | 0.1045 | 0.0091 | 0.3431 | -0.125 | 0.1414 | 0.0009 |
| 2020-03-23 | -0.0239 | -0.125 | 0.1038 | 0.3297 | 0.1331 | -0.125 | 0.1410 | 0.0670 |
| 2020-03-24 | 0.0154 | -0.125 | 0.0972 | 0.1486 | 0.1926 | -0.125 | 0.1370 | 0.0204 |

| Date | | | | | | | |
|---|---|---|---|---|---|---|---|
| 2020-03-25 | 0.1391 | -0.125 | 0.0922 | 0.0042 | 0.2010 | -0.125 | 0.1341 0.0150 |
| 2020-03-26 | -0.0423 | -0.125 | 0.1557 | 0.5952 | 0.4060 | -0.125 | 0.1764 0.0026 |
| 2020-03-27 | -0.0160 | -0.125 | 0.1511 | 0.4704 | 0.1389 | -0.125 | 0.1729 0.1271 |
| 2020-03-28 | 0.0354 | -0.125 | 0.1066 | 0.1325 | -0.3842 | -0.125 | 0.1427 0.0693 |
| 2020-03-29 | 0.1368 | -0.125 | 0.1191 | 0.0279 | 0.1585 | -0.125 | 0.1507 0.0600 |
| 2020-03-30 | 0.1657 | -0.125 | 0.1535 | 0.0582 | 0.0249 | -0.125 | 0.1747 0.3909 |
| 2020-03-31 | 0.3086 | -0.125 | 0.1543 | 0.0050 | -0.2770 | -0.125 | 0.1753 0.3860 |
| 2020-04-01 | 0.0754 | -0.125 | 0.1547 | 0.1951 | -0.2521 | -0.125 | 0.1756 0.4692 |
| 2020-04-02 | 0.1291 | -0.125 | 0.1392 | 0.0680 | 0.3122 | -0.125 | 0.1645 0.0078 |
| 2020-04-03 | 0.2554 | -0.125 | 0.0786 | 0.0000 | -0.0291 | -0.125 | 0.1267 0.4489 |
| 2020-04-04 | 0.2135 | -0.125 | 0.0516 | 0.0000 | 0.0434 | -0.125 | 0.1145 0.1416 |
| 2020-04-05 | 0.3794 | -0.125 | 0.1024 | 0.0000 | 0.1187 | -0.125 | 0.1401 0.0820 |
| 2020-04-06 | 0.0592 | -0.125 | 0.1508 | 0.2220 | 0.4879 | -0.125 | 0.1728 0.0004 |
| 2020-04-07 | 0.0314 | -0.125 | 0.1311 | 0.2327 | -0.2209 | -0.125 | 0.1588 0.5459 |
| 2020-04-08 | 0.0462 | -0.125 | 0.1298 | 0.1871 | -0.0296 | -0.125 | 0.1579 0.5457 |
| 2020-04-09 | 0.5303 | -0.125 | 0.1541 | 0.0000 | 0.0508 | -0.125 | 0.1752 0.3156 |
| 2020-04-10 | 0.2335 | -0.125 | 0.1193 | 0.0027 | 0.0532 | -0.125 | 0.1508 0.2375 |
| 2020-04-11 | 0.0711 | -0.125 | 0.1541 | 0.2033 | -0.0330 | -0.125 | 0.1752 0.5994 |
| 2020-04-12 | 0.2282 | -0.125 | 0.1350 | 0.0089 | 0.0309 | -0.125 | 0.1615 0.3344 |
| 2020-04-13 | 0.2815 | -0.125 | 0.1317 | 0.0020 | -0.1243 | -0.125 | 0.1592 0.9964 |
| 2020-04-14 | 0.2671 | -0.125 | 0.0888 | 0.0000 | -0.3160 | -0.125 | 0.1322 0.1485 |
| 2020-04-15 | 0.2908 | -0.125 | 0.1451 | 0.0042 | -0.2614 | -0.125 | 0.1686 0.4187 |
| 2020-04-16 | 0.1720 | -0.125 | 0.1559 | 0.0567 | -0.1625 | -0.125 | 0.1764 0.8318 |
| 2020-04-17 | 0.2669 | -0.125 | 0.1509 | 0.0094 | 0.0159 | -0.125 | 0.1728 0.4148 |
| 2020-04-18 | 0.2795 | -0.125 | 0.0819 | 0.0000 | -0.0538 | -0.125 | 0.1284 0.5790 |
| 2020-04-19 | 0.2374 | -0.125 | 0.1471 | 0.0138 | -0.1652 | -0.125 | 0.1701 0.8133 |
| 2020-04-20 | 0.3884 | -0.125 | 0.0683 | 0.0000 | 0.0302 | -0.125 | 0.1216 0.2019 |
| 2020-04-21 | 0.4657 | -0.125 | 0.1076 | 0.0000 | -0.1123 | -0.125 | 0.1433 0.9293 |
| 2020-04-22 | 0.3351 | -0.125 | 0.1408 | 0.0011 | 0.0050 | -0.125 | 0.1656 0.4323 |
| 2020-04-23 | 0.3423 | -0.125 | 0.0590 | 0.0000 | -0.0430 | -0.125 | 0.1175 0.4852 |
| 2020-04-24 | 0.4102 | -0.125 | 0.1039 | 0.0000 | -0.1086 | -0.125 | 0.1410 0.9074 |
| 2020-04-25 | 0.2856 | -0.125 | 0.0359 | 0.0000 | -0.0577 | -0.125 | 0.1094 0.5385 |

We also calculated the corresponding p-values of Moran's test given by Table 3 over the time, shown in Figure 7.

Comparing the corresponding p-values of Moran's test, some deviation exist in the statistical timeliness in main cities of provinces from March 21, 2020 to April 25 so it is inevitable that bias occurs in the results. April 20 to 25 is not significantly different in terms of adjacency to main cities. This means that the prevalence of COVID-19 based on relative counts, is the similar in the main cities.

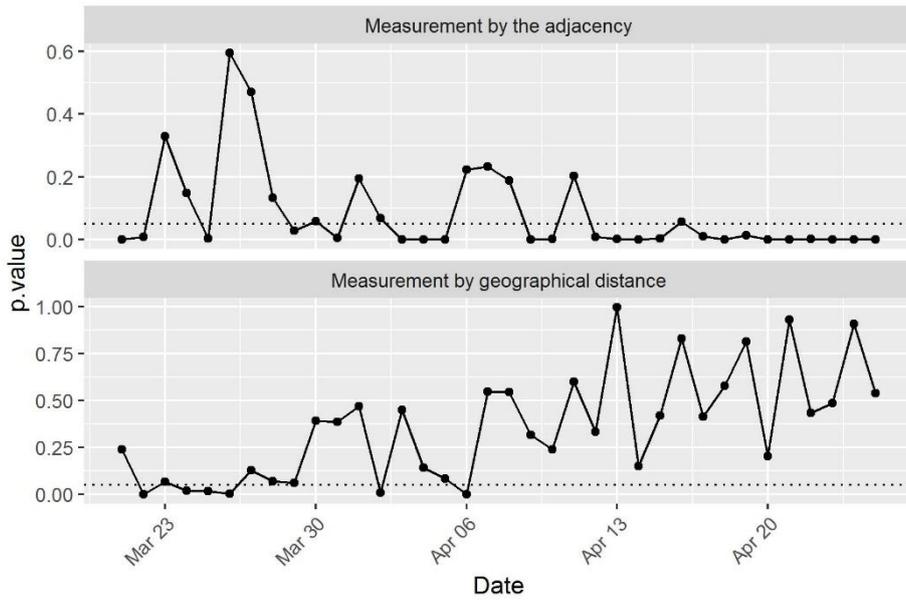

**Figure 7.** The corresponding p-values of Moran's test given by Table 4 over the time, according to the relative counts (absolute counts divided by 1M residents).

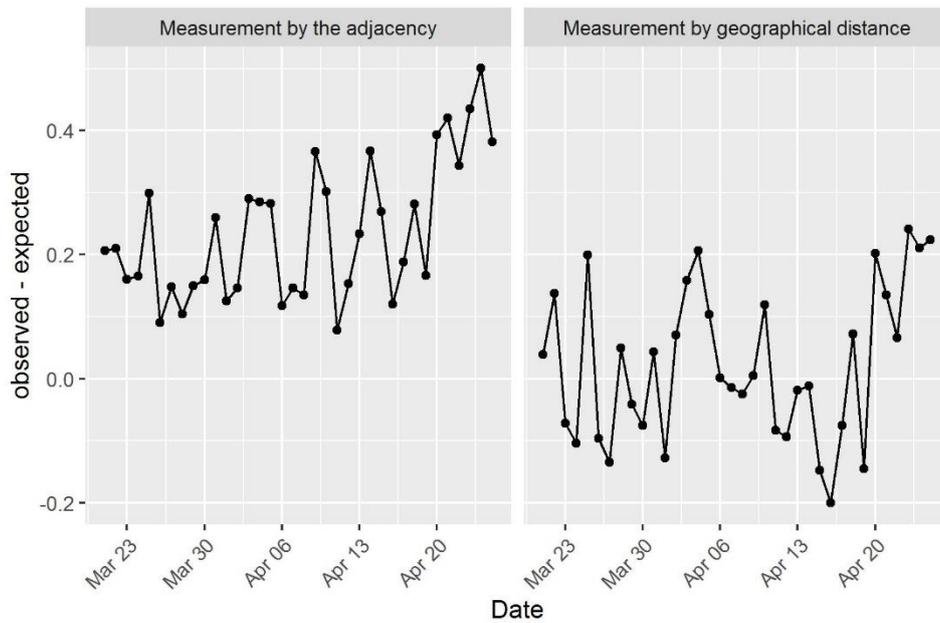

**Figure 8.** The difference between Moran's observed and expected values given by Table 2 over the time, according to the relative counts.

Figure 9 displays observed cumulative confirmed cases, the fitted logistic growth model given by Eq. (2) with v=1 and the corresponding 95% confidence interval for each province. However, the model in Eq. (2) was not fitted on the data associated with the WC province. As it is seen from Figure 9, the cumulative number of confirmed cases in all provinces is described very well by a logistic growth model. The high values of the R-Squared, shown at the bottom-right of each figure, also confirms the goodness of fit of all 8 models, with FS having the lowest R-Square (0.939). Hence, we can rely on predictions given as red lines until June 5 proportionally to the magnitude of the given R-Squared for each province.

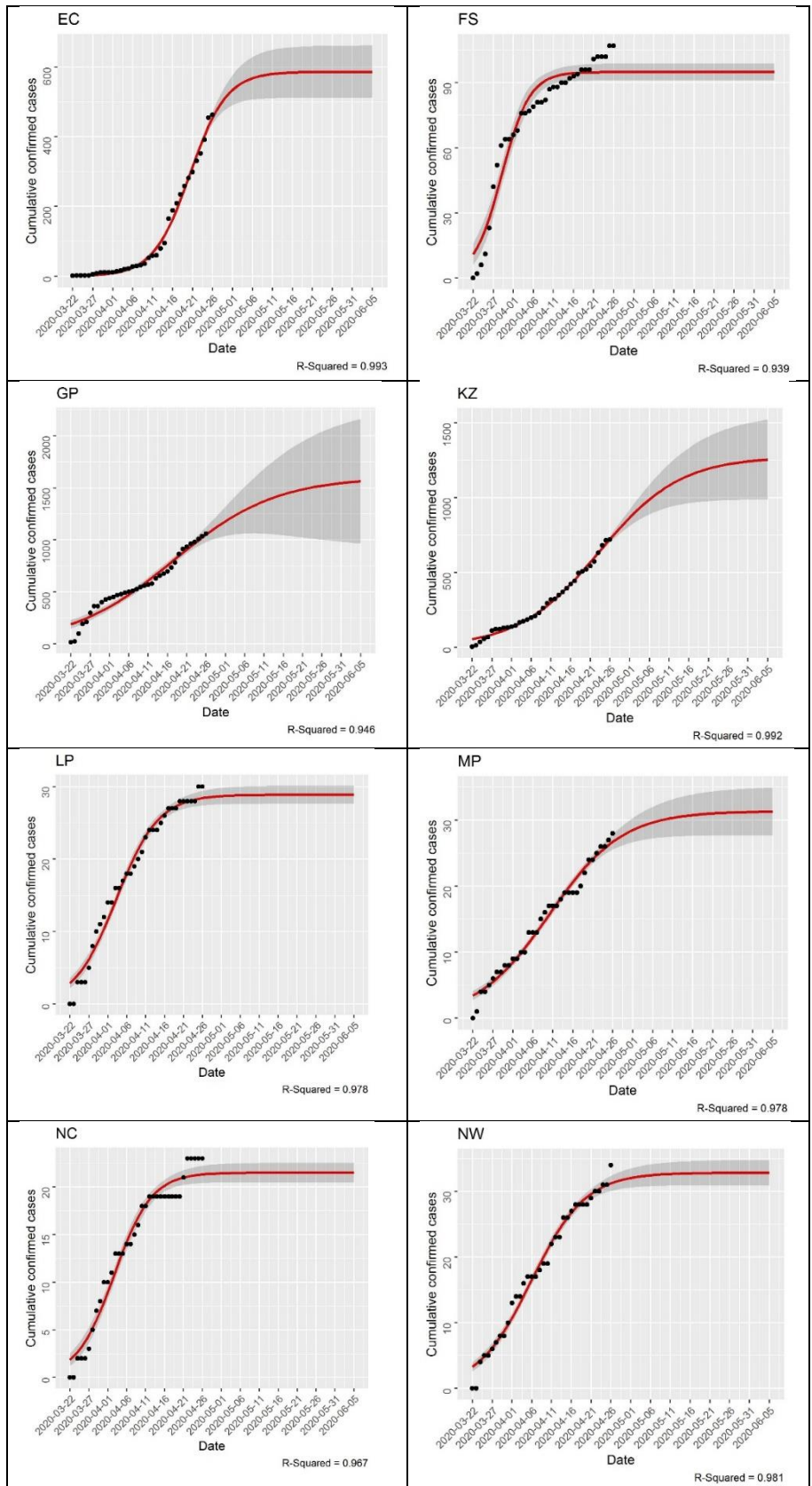

**Figure 9.** Observed cumulative confirmed cases (the black points), the fitted logistic growth model (the red line) given by (2) and the corresponding 95% confidence interval for EC province.

## 5. Conclusions

Despite the inaccuracies associated with medical predictions, identifying hot spots and logistic modelling is still invaluable for better understanding of the spread in South Africa. The results of the Moran index showed the impact of President Cyril Ramaphosa's decision in containing the outbreak by strict lockdown regulations and the inter provisional travelling prohibition has a positive role in tapering the counts. The results indicated that the spreading direction in South Africa is mainly based on adjacent areas to neighbours, and doesn't matter how far the distance is to the infectious centre.

The logistic growth models show a good fit to the provincial data, with R-Square values above 0.9. Visually however it is clear that for certain provinces a different modelling strategy could yield even better results ([36]). These provinces are GP, FS and NC and likely also WC.

With South Africa phasing out the lockdown as of the beginning of May, implementing the risk-adjusted strategy and economic activity plan, South African will be seeing workers returning to their workplace and COVID-19 cases are expected to increase. This initial study highlights the importance of continued analysis and showcases the valuable input that can be obtained from these analyses results.

**Author Contributions:** Conceptualization, M. A., A. B.; Methodology, M. A., A. B.; Software, M. S.; Formal Analysis, M. A., A. B., M. S, M. G., T. C., S. M., B. E.; Data Curation, S. M. Writing – Original Draft Preparation, M. A., A. B., M. S, M. G., T. C., S. M.; Writing – Review & Editing, M. A., A. B., M. S, M. G., T. C., S. M., B. E.; Funding Acquisition, M. A., A. B., S. M., B. E; Validation, M. A., A. B., M. S, M. G., T. C., S. M., B. E.; Investigation, M. A., A. B., M. S, M. G., T. C., S. M., B. E.

**Funding:** This work was based upon research supported by the South African National Research Foundation SARChI Research Chair in Computational and Methodological Statistics (UID: 71199), SRUG190308422768 grant No. 120839; STATOMET at the Department of Statistics at the University of Pretoria, and the Dean's Office at the Faculty of Natural and Agricultural Sciences at the University of Pretoria.

**Conflicts of Interest:** The authors declare no conflict of interest.